\documentclass[aps,pre,twocolumn,groupedaddress,showpacs]{revtex4-2}
\newcommand{\bec}[1]{\mbox{\boldmath $ #1$}}
\usepackage{bm}
\usepackage{graphicx}
\usepackage{color}

\newcommand{\meanrho}{\overline{\rho}}

{}
\newcommand{\meanUU}{\overline{\bm{U}}}

{}
{}

\newcommand{\meanN}{\overline{n}}
\newcommand{\meanP}{\overline{P}}
\newcommand{\meanT}{\overline{T}}

\begin{document}
\title{Experimental study of turbulent transport of nanoparticles in convective turbulence}
\author{I. Shimberg}

\author{O. Shriki}

\author{O.~Shildkrot}

\author{N. Kleeorin}
\author{A. Levy}
\author{I.~Rogachevskii}

\bigskip
\affiliation{
The Pearlstone Center for Aeronautical Engineering
Studies, Department of Mechanical Engineering,
Ben-Gurion University of the Negev, P.O.Box 653,
Beer-Sheva 8410530,  Israel}

\date{\today}
\begin{abstract}
We perform experimental study of turbulent transport of nanoparticles in convective turbulence
with the Rayleigh number $\sim 10^8$ in the air flow.
We measure temperature field in many locations by a temperature probe equipped with 11 E-thermocouples.
Nanoparticles of the size $\sim 70$ nm in diameter are produced by Advanced Electrospray Aerosol Generator.
To determine the number density of nanoparticles, we use Condensation Particle Counter.
We demonstrate that the joint action of turbulent effects (which are important in the core flow) and molecular effects
(which are essential near the boundaries of the chamber) results in an effective accumulation
of nanoparticles at the cold wall of the chamber.
The turbulent effects are characterised by turbulent diffusion and turbulent thermal diffusion of nanoparticles,
while the molecular effects are described by the Brownian diffusion and
thermophoresis, as well as the adhesion of nanoparticles at the cold wall of the chamber.
In different experiments in convective turbulence in a chamber with
the temperature difference $\Delta T$ between the bottom and top walls  varying
between $\Delta T= 29$ K to $\Delta T= 61$ K,
we find that the mean number density of nanoparticles decreases exponentially in time.
For instance, the characteristic decay time of the mean number density of nanoparticles varies
from 12.8 min for $\Delta T= 61$ K to 24 min for $\Delta T= 29$ K.
For better understanding of experimental results, we
perform  one-dimensional mean-field numerical simulations of the evolution of the mean number density of nanoparticles
for conditions pertinent to the laboratory experiments.
The obtained numerical results  are in a good agreement with the experimental results.
\end{abstract}

\maketitle

\section{Introduction}
\label{sect1}

Turbulence \cite{MY71,MY75,MC90,F95,P2000,LE08,DA13}
and the associated turbulent transport of particles \cite{CSA80,ZRS90,BLA97,SP2006,ZA08,CST11,RI21}
have been investigated systematically for more than a hundred years
in theoretical, experimental and numerical
studies. But some fundamental questions remain.
It is well-known that turbulence
results in a sharp increase of the effective diffusion coefficient \cite{TA1922}.
In addition to turbulent diffusion, there are various turbulent effects resulting
in formation of inhomogeneities in spatial distribution of particles.

In a non-stratified inhomogeneous turbulence, turbophoresis of particles
due to a combined effect of particle inertia and inhomogeneity of turbulence, can occur
\cite{CTT75,R83,G97,EKR98,G08,MHR17}.
Turbophoresis causes an appearance of the additional non-diffusive
flux of inertial particles proportional to the mean particle velocity,
${\bm V}_{\rm turbo} \propto - f_{\rm turbo}({\rm St},{\rm Re}) \, \bec{\nabla}
\langle {\bm u} \rangle^2$, where ${\bm u}$ is the turbulent fluid
velocity, ${\rm St} = \tau_{\rm p}/\tau_\eta$ is the Stokes number, $\tau_\eta=\tau_0/{\rm Re}^{1/2}$
is the Kolmogorov viscous time, $\tau_{\rm p} = m_{\rm p} / (3 \pi \rho \, \nu d)$ is the Stokes time for the small
spherical particles of the diameter $d$ and mass $m_{\rm p}$,  and $\rho$
is the fluid density,
${\rm Re}= \ell_0\, u_0/\nu$ is the fluid Reynolds number,
$u_0$ is the characteristic turbulent velocity at the integral scale $\ell_0$ of turbulent motions
and $\nu$ is the kinematic fluid viscosity. Due to turbophoresis
inertial particles are accumulated in the vicinity of the minimum of the turbulent
intensity.

In a turbulent flow with a non-zero mean temperature gradient an additional
turbulent flux of particles appears in the direction opposite to that of
the mean temperature gradient \cite{EKR96,EKR97}.
This phenomenon of turbulent thermal
diffusion causes a non-diffusive turbulent flux of particles in the direction
of the turbulent heat flux (opposite to the mean temperature gradient).
Turbulent thermal diffusion results in accumulation of particles
in the vicinity of the mean temperature minimum which leads to the formation of
inhomogeneous spatial distributions of the mean particle number density.
Turbulent thermal diffusion has been intensively investigated analytically
\cite{EKRS00,EKRS01,PM02,RE05,AEKR17,RKB18}
using different theoretical approaches.
This phenomenon has been detected in laboratory experiments for micron-size particles in air flows
in oscillating grid turbulence \cite{BEE04,EEKR04,EEKMR06}
and in multi-fan produced turbulence  \cite{EEKR06}.
Turbulent thermal diffusion has been also detected in direct numerical
simulations \cite{HKRB12} and observed in the atmospheric turbulence \cite{EKR00,SSEKR09}.
This phenomenon has been also shown to be important in astrophysical
turbulent flows \cite{H16}.

In spite of intensive laboratory studies of turbulent thermal
diffusion of micron-size particles,
this effect has not yet been investigated for nanoparticles.
To study turbulent thermal diffusion of nanoparticles, one need to take into account the dependence
of the characteristic relaxation time of particles (the Stokes time) on the Knudsen number, ${\rm Kn}=2\lambda/d$ \cite{SP2006},
where $\lambda$ is the mean free path of molecules and $d$ is the particle diameter.
Note also that various aspects of transport of nanoparticles in turbulent fluid flows have been studied
in a number of papers
\cite{E04,WF05,W07,DW07,KP09,YL10,LC15,RA16,VG16}.

In the present paper, we discuss results of the experimental study of turbulent
transport of nanoparticles in convective turbulence.
The effect of turbulent thermal diffusion in the core flow and
molecular thermophoresis nearby the cold boundary are main effects
resulting in accumulation of nanoparticles nearby the cold wall.
Our main goal is to understand how fast  it is possible
to accumulate nanoparticles nearby the cold wall of the chamber, and how to control
this process.
The effect of nanoparticle accumulation depends
on properties of the temperature stratified
turbulence.
To understand how to control nanoparticle accumulation,
we perform measurements of  temperature fields in many locations
in the chamber and conduct the direct measurements of
the number density of nanoparticles in the chamber.
This allows us to determine the time-dependence of the mean number density
of nanoparticles in the chamber.
To understand the experimental results, we also
perform  one-dimensional mean-field numerical simulations of transport of nanoparticles
for conditions pertinent to the laboratory experiments.

This paper is organized as follows.
In Section~\ref{sect2} we discuss the physics of turbulent thermal diffusion and
in Section~\ref{sect3} we outline
the numerical setup for the  mean-field numerical simulations of turbulent transport of nanoparticles.
In Section ~\ref{sect4} we describe the experimental setup and measurements techniques.
In Section ~\ref{sect5} we discuss the experimental results and compare these results
with those of the mean-field numerical simulations performed for conditions pertinent to the
laboratory experiments.
Finally, conclusions are drawn in Section~\ref{sect6}.

\section{Turbulent thermal diffusion}
\label{sect2}

In this Section we discuss the physics of the phenomenon of turbulent thermal
diffusion.
The evolution of the number density $n(t,{\bm r})$ of small particles  in a turbulent flow
is determined by the convective diffusion equation,
\begin{eqnarray}
{\partial n \over \partial t} + {\bm \nabla} \cdot
(n \, {\bm U}_{\rm p}) = - {\bm \nabla}  \cdot {\bm J}_{\rm M},
\label{P1}
\end{eqnarray}
where ${\bm U}_{\rm p}$ is the particle velocity advected by a turbulent
temperature stratified fluid flow,
${\bm J}_{\rm M}$ is the molecular flux of
particles that is given by
\begin{eqnarray}
{\bm J}_{\rm M} = - D \left[{\bm \nabla} n + k_{\rm t} {{\bm \nabla} T
\over T} + k_{\rm p} {{\bm \nabla} P
\over P} \right] ,
\label{P2}
\end{eqnarray}
where $T$ and $P$ are the temperature and pressure of the surrounding fluid,
respectively.
The first term  ($\propto {\bm \nabla} n$) in the right-hand side of Eq.~(\ref{P2}) for the molecular
flux of particles, describes Brownian (molecular) diffusion of particles, the second term accounts for the molecular flux of
particles which is driven by the fluid temperature gradient
${\bm \nabla} T$ (thermophoresis for
particles or molecular thermal diffusion for gases), and the third term determines the
molecular flux of particles which is driven by the fluid pressure gradient
${\bm \nabla} P$ (molecular barodiffusion). Here $D$ is
the coefficient of the Brownian diffusion, $k_{\rm t} \propto n$ is the
thermal diffusion ratio and $D k_{\rm t}$ is the  coefficient of
thermal diffusion, $k_{\rm p}\propto n$ is the barodiffusion ratio
and $D k_{\rm p}$ is the  coefficient of barodiffusion.

In a turbulent flow, large-scale dynamics of particles
is determined by the equation for the mean particle number density $\meanN(t,{\bm r})$:
\begin{eqnarray}
{\partial \meanN \over \partial t} + {\bm \nabla} \cdot
(\meanN \, \meanUU_{\rm p}) = - {\bm \nabla} \cdot (\overline{\bm J}_{\rm T} + \overline{\bm J}_{\rm M}),
\label{P3}
\end{eqnarray}
where $\meanUU_{\rm p}$ is the mean particle velocity,
$\overline{\bm J}_{\rm M}$ is the averaged molecular flux of particles,
\begin{eqnarray}
\overline{\bm J}_{\rm M} = - D \, \left [{\bm \nabla} \meanN + k_{\rm t} \, {{\bm \nabla} \meanT
\over \meanT} + k_{\rm p} \, {{\bm \nabla} \meanP \over \meanP} \right],
\label{P4}
\end{eqnarray}
$\meanT$ and $\meanP$ are the mean fluid temperature and pressure, respectively,
$\overline{\bm J}_{\rm T}$ is the turbulent flux of particles \cite{EKR96,EKR97},
\begin{eqnarray}
\overline{\bm J}_{\rm T} = - D_{\rm T} \, \left [{\bm \nabla} \meanN + k_{\rm T} \, {{\bm \nabla} \meanT
\over \meanT} + k_{\rm P} \, {{\bm \nabla} \meanP \over \meanP} \right] .
\label{P5}
\end{eqnarray}
Here $D_{\rm T}$  is the coefficient of turbulent diffusion,
$k_{\rm T} = \meanN \, f_{\rm p}(d, {\rm Re}, \meanT)$
can be interpreted as the turbulent thermal diffusion ratio and
$D_{\rm T} k_{\rm T}$ is the coefficient of turbulent thermal diffusion,
$k_{\rm P} = - \meanN$ can be interpreted as the turbulent barodiffusion
ratio and $D_{\rm T} k_{\rm P}$ is the coefficient of turbulent
barodiffusion, $f_{\rm p}(d, {\rm Re}, \meanT)$ is the function
which depends on the particle diameter, the Reynolds number and the
mean  fluid temperature \cite{AEKR17}.

The physics of the effect of turbulent thermal diffusion
for solid particles $\rho_{\rm p} \gg \rho$ is as follows \cite{EKR96,EKR97},
where $\rho_{\rm p}$ is the material density of the particles.
The inertia causes particles inside the turbulent eddies to drift
out to the boundary regions between eddies due to the centrifugal inertial force.
Indeed, for large P\'eclet numbers, when molecular diffusion of
particles in Eq.~(\ref{P1}) can be neglected, we obtain that
$\bec\nabla {\bf \cdot} \, {\bm U}_{\rm p} \approx  - n^{-1} \, {\rm d}n / {\rm d}t$.
On the other hand, for inertial particles, $\bec\nabla {\bf \cdot} \, {\bm U}_{\rm p}
= (\tau_{\rm p} / \rho)  \,\bec\nabla^2 P$ \cite{EKR96}. Therefore, in
regions with maximum fluid pressure fluctuations (where
$\bec\nabla^2 p < 0)$, there is accumulation of
inertial particles, i.e.,  ${\rm d} n' / {\rm d}t
\propto - \overline{n} \, (\tau_p /\overline{\rho}) \,\bec\nabla^2 p
> 0$.
These regions have low vorticity and high
strain rate.
Here $n'$ and $p$ are fluctuations of particle number density and fluid pressure, respectively,
and $\meanrho$  is the mean density of the surrounding
fluid.
Similarly, there is an outflow of inertial
particles from regions with minimum fluid
pressure fluctuations.

In homogeneous and isotropic turbulence with a zero gradient
of the mean temperature, there is no preferential direction,
so that there is no large-scale effect of particle accumulation,
and the pressure (temperature) of
the surrounding fluid is not correlated with the turbulent velocity field. The only
non-zero correlation is $\langle({\bm u} \cdot {\bm \nabla})p\rangle$,
which contributes to the flux of the turbulent
kinetic energy density.

In temperature-stratified turbulence, fluctuations of fluid temperature $\theta$
and velocity ${\bm u}$ are correlated due to a
non-zero turbulent heat flux, $\langle
\theta \, {\bm u} \rangle\not=\bm{0}$. Fluctuations of
temperature cause pressure fluctuations, which
result in fluctuations of the number density of
particles.
Note that in the mechanism of turbulent thermal diffusion, only pressure fluctuations which
are correlated with velocity fluctuations
due to a non-zero turbulent heat flux play a crucial role.
Increase of the fluid pressure fluctuations
is accompanied by an accumulation of particles,
and the direction of the mean flux of particles
coincides with that of the turbulent heat flux.
The turbulent flux of particles is directed to the
minimum of the mean temperature, and the
particles tend to be accumulated in this
region \cite{RI21}.

The similar effect of accumulation of particles
in the vicinity of the mean temperature minimum and the formation of
inhomogeneous spatial distributions of the mean particle number density
exists also for non-inertial particles or gaseous admixtures
in turbulent compressible flows \cite{EKR97,HKRB12,RKB18}.
Compressibility in a low-Mach-number stratified turbulent fluid flow
[with ${\bm \nabla} \cdot {\bm u}
\approx - ({\bm u} \cdot {\bm \nabla} \meanrho) / \meanrho
\not= 0] $ causes an additional non-diffusive component of the turbulent flux of non-inertial
particles or gases, and results in the formation of large-scale inhomogeneous structures
in spatial distributions of non-inertial particles.
In a temperature stratified turbulence, preferential concentration of particles
caused by turbulent thermal diffusion
can occur in the vicinity of the minimum of the mean temperature. The latter effect is caused by the
compressibility of the turbulent fluid flow
$ [{\bm \nabla} \cdot
{\bm u} \propto ({\bm u} \cdot {\bm \nabla} \meanT) / \, \meanT \not= 0]$.
This effect plays an important role in the dynamics of gaseous pollutants
in the stratified atmospheric turbulence.

\section{Setup for mean-field numerical simulations}
\label{sect3}

For better understanding of the experimental results obtained in this study, we perform one-dimensional mean-field numerical simulations  of turbulent transport of nanoparticles for the conditions pertinent to the laboratory experiments. In particular, we study the evolution of the mean number density of nanoparticles, $\meanN(t,z)$, by solving numerically the evolutionary equation for the mean number density of nanoparticles:
\begin{eqnarray}
{\partial \meanN \over \partial t} + \nabla_z \left[\left(V_z^{\rm (eff)}+V_z^{\rm (tp)}\right) \, \meanN - (D+D_{\rm T}) \,\nabla_z \meanN \right]=0 .
 \nonumber\\
 \label{IC2}
\end{eqnarray}
This equation takes into account the total transport effective velocity ${\bm V}^{\rm (eff)}+{\bm V}^{\rm (tp)}$ due to turbulent thermal diffusion and thermophoresis, where ${\bm V}^{\rm (eff)}$ is the effective pumping velocity caused by turbulent thermal diffusion
and ${\bm V}^{\rm (tp)}$ is the thermophoretic velocity (see below). This equation also takes into account the total particle diffusion $D+D_{\rm T}$, where $D$ is the coefficient of the Brownian diffusion, $D_{\rm T} = u_0 \, \ell_0 / 3$  is the coefficient of turbulent diffusion and $u_0$ is the characteristic turbulent velocity at the integral turbulence scale $\ell_0$.
In the one-dimensional mean-field numerical simulations,
we have not taken into account the mean velocity field of the large-scale circulation
produced in a small-scale convective turbulence.

Turbulent thermal diffusion is described in terms of  the effective pumping velocity
${\bm V}^{\rm (eff)}$ resulting in non-diffusive turbulent particle flux, $\meanN \, {\bm V}^{\rm (eff)} $.
The effective pumping velocity caused by turbulent thermal diffusion of nanoparticles is given by
\begin{eqnarray}
{\bm V}^{\rm (eff)} = - D_T \, \alpha(d, {\rm Kn}, {\rm Re}) \, {{\bm \nabla} \meanT \over \meanT} ,
\label{BBB8}
\end{eqnarray}
where
\begin{eqnarray}
 \alpha(d, {\rm Kn}, {\rm Re}, \meanT) =1 + {\tau_p(d, {\rm Kn})  \over \tau_0} \, {\rm Re}^{1/4} \, \ln({\rm Re})  \, \left({L_{\rm eff} \over \ell_0}\right) ,
 \nonumber\\
\label{BBB9}
\end{eqnarray}
(see \cite{EKR98,EKR00,EKR13}), where
$L_{\rm eff} = 2 c_s^2 \tau_\eta^{3/2}/3 \nu^{1/2}$ is the effective length scale, $c_s$ is the
sound speed,  $\tau_{\rm p} = m_{\rm p} \, C_c / (3 \pi \rho \, \nu d)$ is the Stokes time
for nanoparticles and $C_c = 1 + {\rm Kn} [1.257 + 0.4 \exp (-1.1 {\rm Kn})]$ is the slip correction
factor \citep{AR82}.
For instance, $C_c=4.95$ for $d=50$ nm, $C_c=3.7$ for $d=70$ nm and $C_c=2.85$ for $d=100$ nm
\citep{SP2006}.

In turbulent flows, at the vicinity of the boundaries
where the intensity of velocity fluctuations drastically decreases,
the molecular effects (e.g, molecular diffusion
and thermophoresis) become more important than the turbulent effects.
The thermophoretic velocity can be estimated as $V^{\rm (tp)} =-  f_{\rm tp} \, \nu \, \bec{\nabla}
\meanT/\meanT$, where $f_{\rm tp}$ is a function of Knudsen number,
the particle size and the ratio of the heat conductivities of the particle and the fluid.
For small particles (i.e for large Knudsen numbers, ${\rm Kn} \gg 1$),
the coefficient $f_{\rm tp}= 3/4$  for mirror rebound
of the gas molecules from the particles and $f_{\rm tp}= 1/2$  for diffuse
evaporation of the molecules when they "forget" the direction and value
of their velocity prior to the impact \cite{DSR66}.
Different aspects related to molecular effects have been
studied in a number of publications
\cite{TC80,CG98,SP02,MC19}.

In the mean-field numerical simulations, we use the following boundary conditions:
\begin{itemize}
\item{
at the bottom boundary ($z=0$), the total flux of particles,
\begin{eqnarray}
F_z^{(n)} = \left(V_z^{\rm (eff)}+V_z^{\rm (tp)}\right) \, \meanN - (D+D_T) \,\nabla_z \meanN,
\label{Z1}
\end{eqnarray}
vanishes;}
\item{
at the upper boundary ($z=L_z$), the vertical gradient of the mean number density of nanoparticles $\nabla_z \meanN$ vanishes.
The latter condition implies that all particles  in the vicinity of the cold wall of the chamber (the upper boundary) are trapped due to adhesion of particles at the wall.}
\end{itemize}

The total transport effective velocity ${\bm V}^{\rm (eff)}+{\bm V}^{\rm (tp)}$ caused by turbulent thermal diffusion and thermophoresis,
depend on the vertical profile of the mean temperature.
We determine semi-analytically the vertical profile of the mean temperature.
To this end, we take into account that
the vertical component of the turbulent heat flux  is $F_z=\langle u_z \, \theta \rangle = - \kappa_{_{T}} \nabla_z \meanT$,
where $\kappa_{_{T}}= \ell_z \, u_z^{\rm (rms)} = (2 \ell_z^4 \beta F_z)^{1/3}$ is the turbulent heat conductivity,
$\beta=g/T_\ast$ is the buoyancy parameter, ${\bf g}$  is the gravity acceleration,
$T_\ast$ is the characteristic temperature in the basic reference state, and
$\ell_z$ is the vertical integral turbulent scale.
We take into account that in convective turbulence, the characteristic vertical turbulent velocity
is $u_z^{\rm (rms)} = (2 \ell_z \beta F_z)^{1/3}$.
The latter equation follows from the steady-state solution
of the budget equation for the turbulent kinetic energy, which allows us to relate
the turbulent velocity with the turbulent heat flux.
Next, we use the following model for the vertical integral length scale:
in the core flow (for $z_\ast  \leq z \leq L_z-z_\ast$), the vertical integral length is $\ell_z=\ell_0$;
and near the bottom and top boundaries where the intensity of turbulence vanishes, the vertical integral length is
$\ell_z=\ell_0 \, \phi(z)$.
Here $L_z$ is the thickness of the convective layer (the vertical height of the
chamber), $z_\ast=\epsilon \, L_z$  with $\epsilon \ll 1$ and $\phi(z)$ is the rapidly decreasing smooth function
that varies from 1 to 0.

At the vicinity of the boundaries the molecular diffusion of the mean temperature is taking into account, so that
the total vertical heat flux $F_z^{\rm (tot)}$ is the sum of turbulent,  $ - \kappa_{_{T}} \nabla_z \meanT$, and molecular, $- \kappa \, \nabla_z \meanT$, vertical heat fluxes, i.e., $F_z^{\rm (tot)}= - (\kappa_{_{T}} + \kappa)\nabla_z \meanT$,
where $\kappa$ is the molecular temperature diffusivity.
The latter equation can be  reduced to a cubic  equation $K M^3 + M^2 - K = 0$ for $M =\left(F_z/F_z^{\rm (tot)}\right)^{1/3}$,
where $K=\kappa^{-1} \, \left(2 \ell_z^4 \beta F_z^{\rm (tot)}\right)^{1/3}  $.
Numerical solution of this cubic equation allows us to obtain the vertical profile for the mean temperature.
This vertical  profile of the mean temperature is used to determine the total transport effective
velocity ${\bm V}^{\rm (eff)}+{\bm V}^{\rm (tp)}$ caused by turbulent thermal diffusion and thermophoresis
in the mean-field numerical simulations
of nanoparticles in convective turbulence  for the conditions pertinent
to the laboratory experiments.
The numerical simulations are performed using the Matlab.
The numerical results will be compared with the experimental results  (see Section~\ref{sect5}).

\section{Experimental setup}
\label{sect4}

\begin{figure}
\centering
\includegraphics[width=7.0cm]{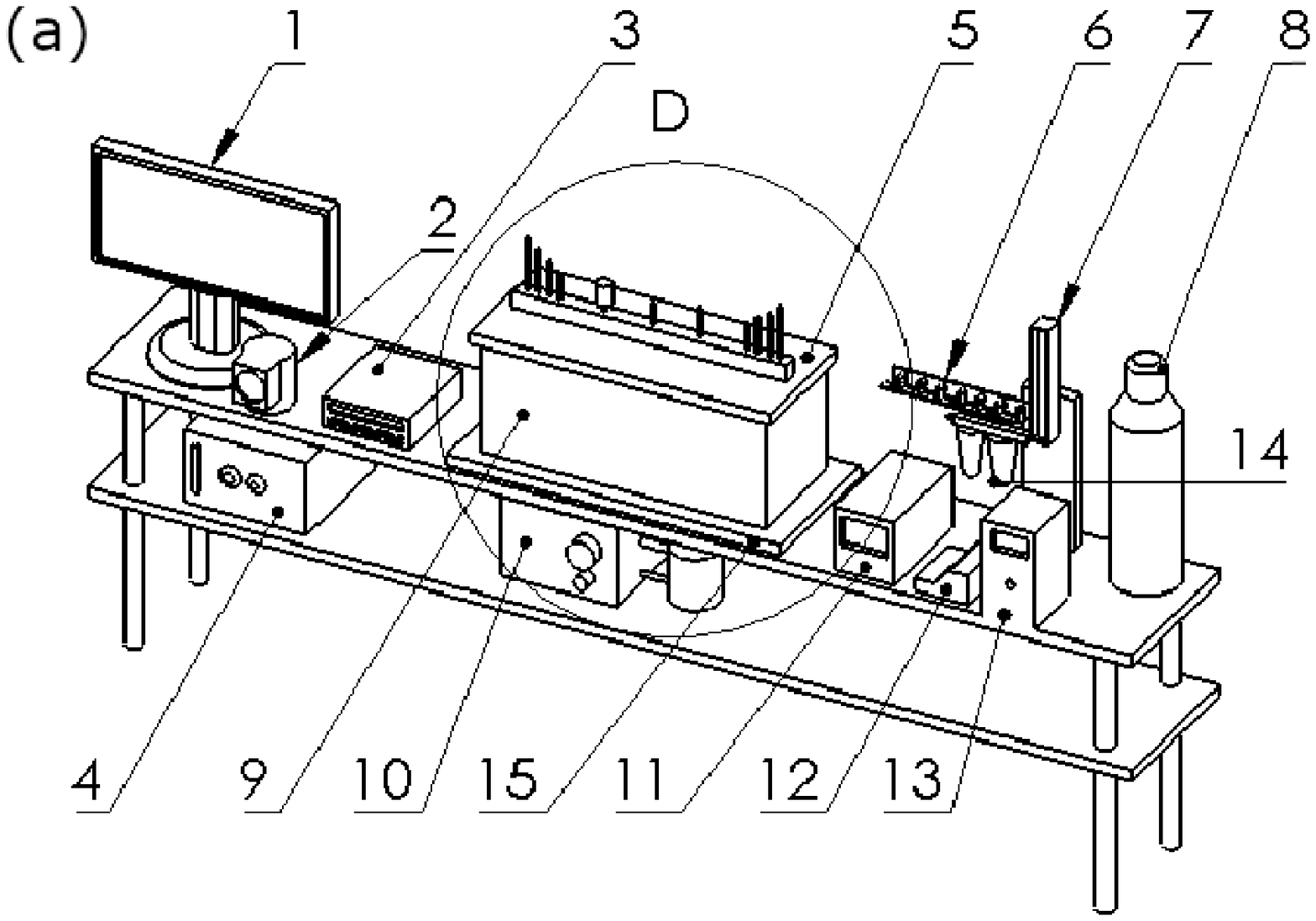}
\includegraphics[width=7.0cm]{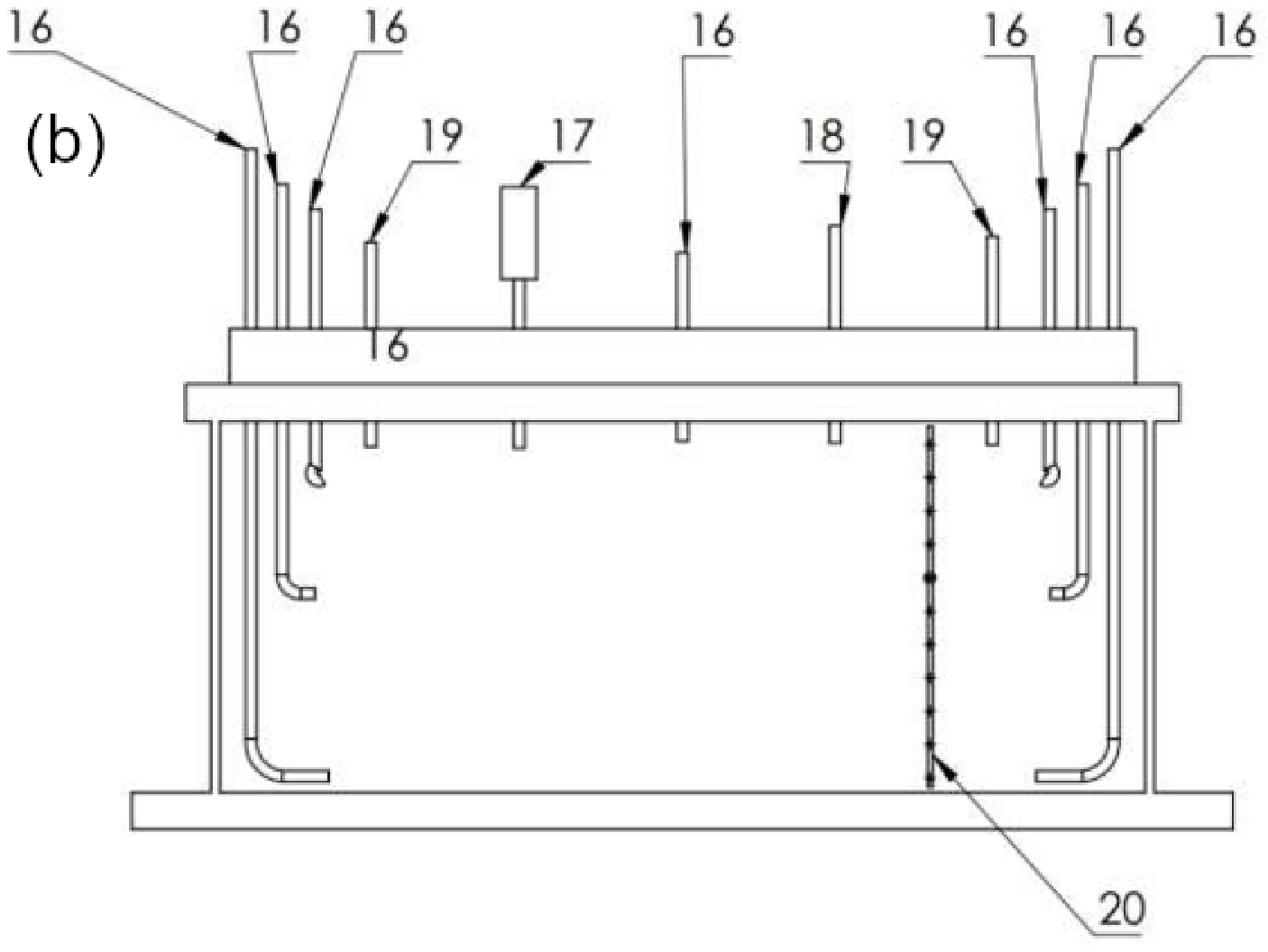}
\caption{\label{Fig1}
A scheme of the experimental set-up:\\
(a). Upper panel: (1) display; (2) pressure measurements device;
(3) data collection for temperature and humidity measurements; (4) computer;
(5) cooler plate; (6-7) controllers for nanoparticle number density measurements;
(8) container for CO$_2$ for nanoparticles generator; (9) chamber (see also the bottom panel);
(10) cooling system; (11) nanoparticles generator; (12) spray for nanoparticles generator;
(13)  condensation particle counter; (14) air quality controller nanoparticles generator;
(15) heating system.\\
(b). Bottom panel: (16) probes for nanoparticle number density measurements;
(17) probe for humidity measurements; (18) probe for pressure measurements;
(19) air supply tube; (20) probes for temperature measurements.}
\end{figure}

\begin{figure}
\centering
\includegraphics[width=7.0cm]{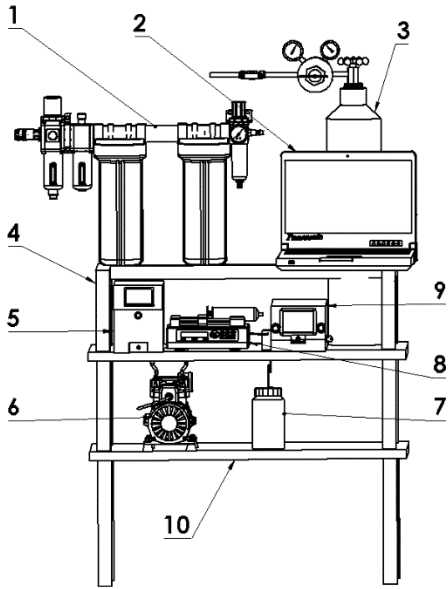}
\caption{\label{Fig2}
A scheme of the experimental set-up:\\
 (1) air filter  and dryer for particles generator;
 (2) display;
 (3) gas cylinder CO$_2$ for particles generator;
 (4) frame;
 (5) particles counter;
 (6) vacuuum pump of particles counter;
 (7) butanol waste container (drain balun of counter);
 (8) particles supply pump;
 (9) particles generator;
 (10) frame.}
\end{figure}

 We conduct various sets of experiments in a
temperature stratified convective turbulence.
The experimental setup with an isolated chamber is designed for measurements of spatial distributions of the fluid temperature
and the number density of nanoparticles in turbulent convection (see Figs.~\ref{Fig1}--\ref{Fig2}).
The chamber is fully isolated using high-quality glass, and there is also control humidity
in the chamber.

The experiments are conducted with air as the working fluid in a rectangular chamber with
dimensions $L_x \times L_y \times L_z$, where $L_x = L_z = 20$ cm, $L_y=50$ cm
and the $z$ axis is in the vertical direction (see Fig.~1, bottom panel).
A vertical mean temperature gradient in the turbulent air flow
is formed by attaching two aluminium heat exchangers to the bottom
and top walls of the test section (a heated bottom and a cooled top wall of the chamber).
A thickness of the massive aluminium heat exchangers is 2 cm. The top plate is a bottom wall of the tank with cooling water. Cold water is pumped into the cooling system through two inlets and flows out through two outlets located at the side wall of the cooling system.
The bottom plate is attached to the electrical heater with wire tightly laid in the grooves milled in the aluminum plate and provided uniform heating. Energy supplied to the heater is varied in order to obtain necessary temperature difference between heater and cooler. Characteristic time of heater is approximately 180 min that stabilize applied temperature during measurements.

The temperature field is measured with  a
temperature probe equipped with 11
E-thermocouples (with the diameter of 0.13 mm and
the sensitivity of $\approx 65 \, \mu$V/K)
attached to a rod with a diameter 4 mm (see Fig.~\ref{Fig1}, bottom panel).
The spacing between thermocouples along the rod
can be changed from 10 to 20 mm. Each thermocouple is inserted into a
1 mm diameter and 45 mm long case. A tip of a
thermocouple protruded at the length of 15 mm out
of the case.
Thermocouples of type {\it E} are used for the temperature measurements
in the core flow, while thermocouples of type {\it K} are used
for temperature measurements at the heater and the cooler.
All thermocouples are built by a manufacturer.
Calibrations of all E-thermocouples in the temperature probe
is performed for three experiments with boiled water
$(T=373$ K), the cold water with ice $(T=273$ K) and
water with intermediate temperature $(T=296$ K).
Comparisons is performed using a precision temperature
measurement manufactured device.

The temperature is measured for 11
rod positions with 10-20 mm intervals in the
horizontal direction.
A sequence of 240 temperature readings for every
thermocouple at every rod position is recorded
and processed.
We measure the temperature field in many locations.
Performing direct continuous measurements of the temperatures
at the cooled top surface and at the
heated bottom surface and using a standard device for supporting constant
temperature difference $\Delta T$ between the top and bottom surfaces
(Contact voltage regulator TDGC-2K),
we control the constant temperature difference
$\Delta T$ during the experiments.

Similar measurement techniques
and data processing procedure have been used by us
previously in the experimental study of different
aspects of turbulent convection and stably stratified turbulence
\cite{BEKR09,EEKR11,EEKR13,BEKR20},
investigations of turbulent thermal
diffusion \cite{EKR96,EKR97,EKRS00,EKRS01}
of micron-size particles in the oscillating grid turbulence
\cite{BEE04,EEKR04,EEKMR06,AEKR17}
and in turbulence
produced by the multi-fan generator \cite{EEKR06},
and investigations of small-scale particle clustering \cite{EKR96a,EKR02,EKR13}
in the oscillating grid turbulence \cite{EKR10}.

To study turbulent thermal diffusion in the experiments
we use 70 nm nanoparticles produced by Advanced Electrospray Aerosol
Generator (Model 3482).
To determine the number density of nanoparticles, we use Condensation
Particle Counter (Model 3750).
We conduct experiments for different temperature differences (from $\Delta T = 29$ K
up to $\Delta T =61$~K) between the top and bottom walls as well as for isothermal flow.
The particle number density measurements
are performed for three heights $z$ and two horizontal coordinates $y$.

\section{Results}
\label{sect5}

\begin{figure}
\centering
\includegraphics[width=8.5cm]{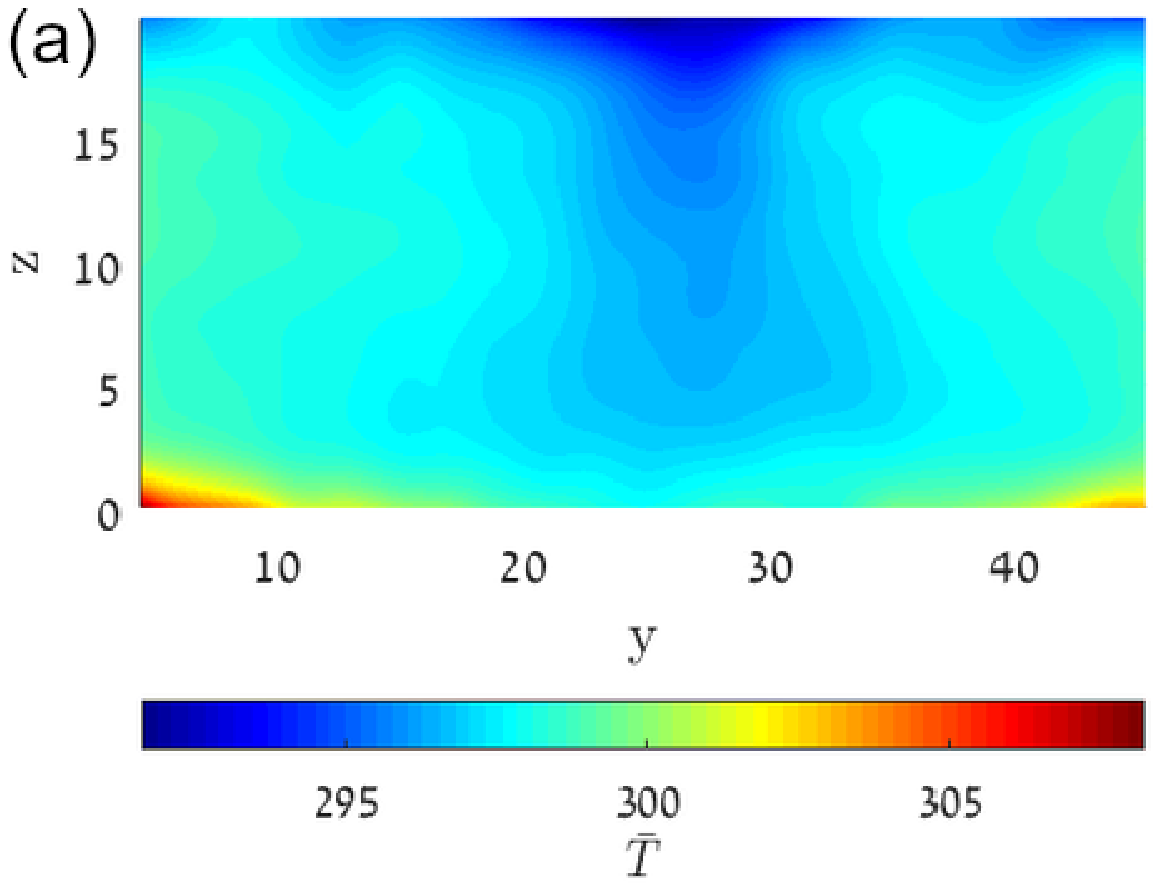}
\includegraphics[width=8.5cm]{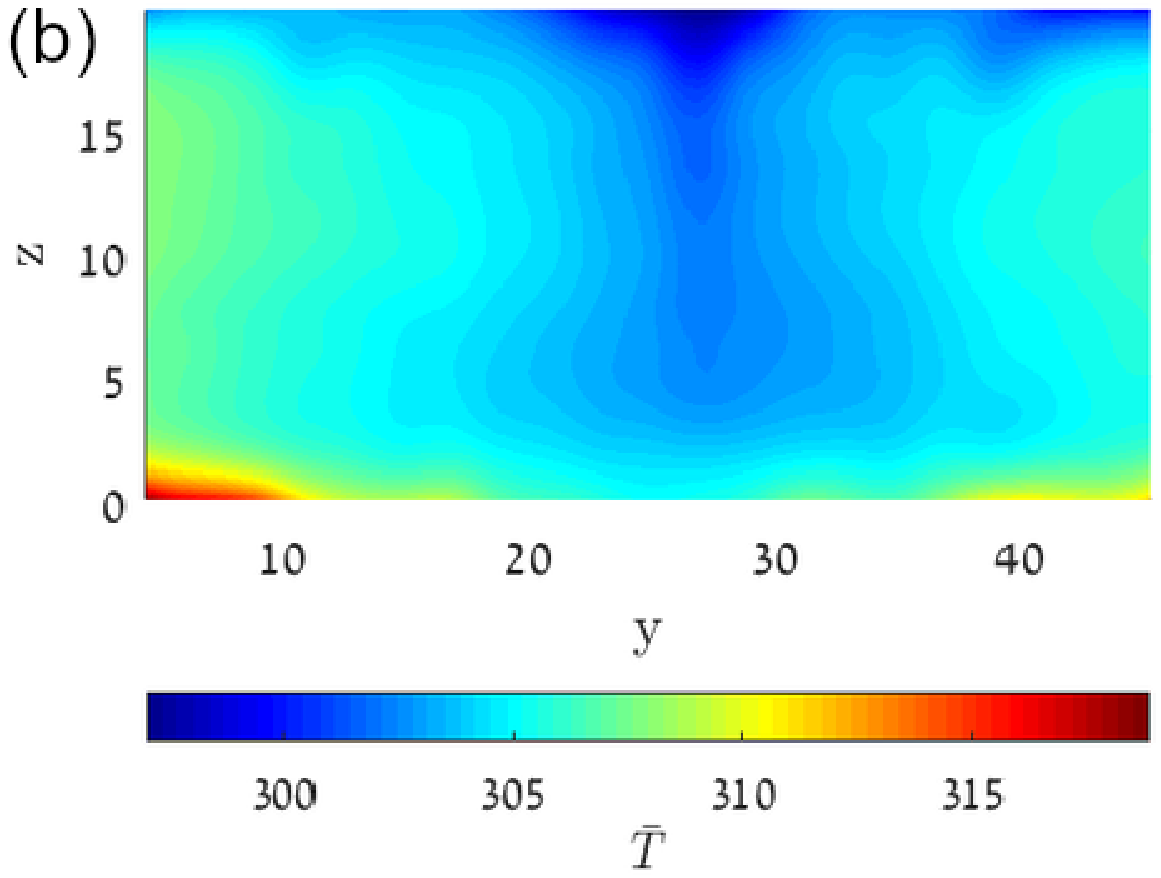}
\includegraphics[width=8.5cm]{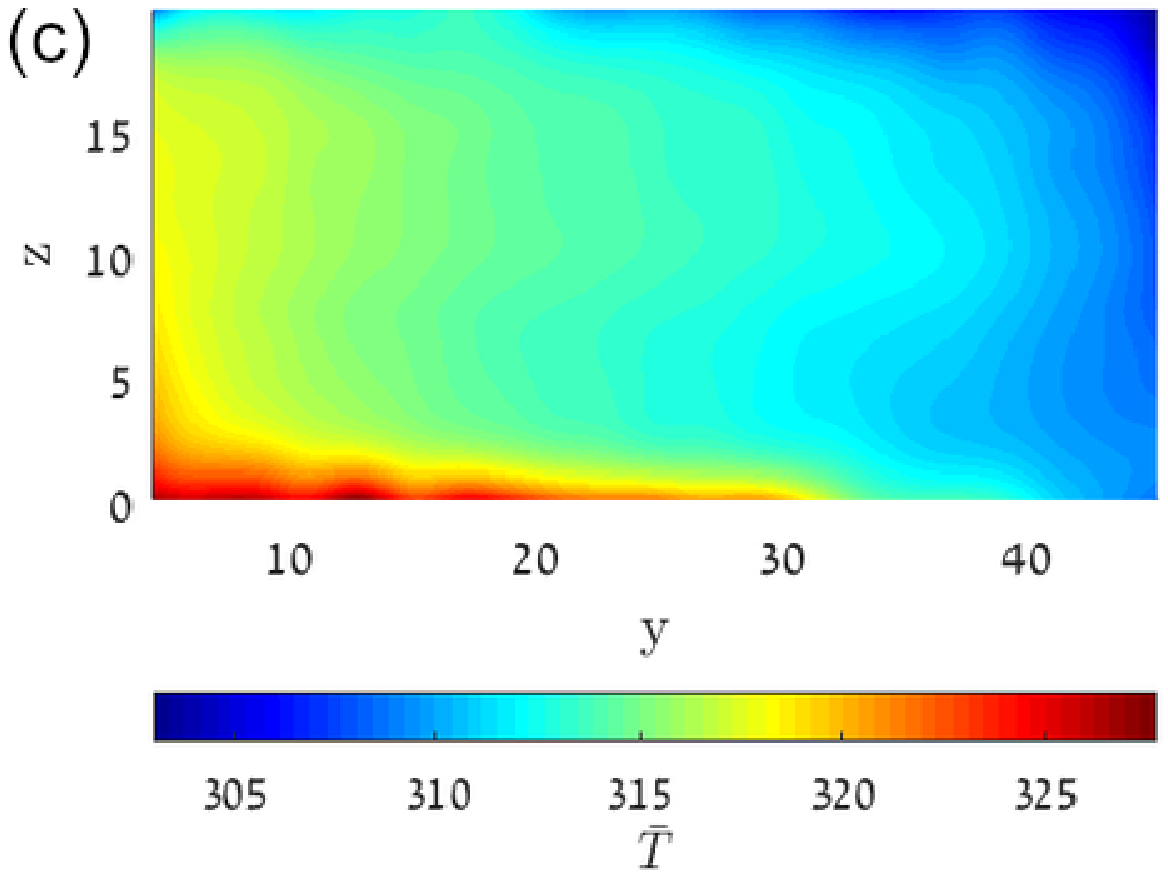}
\caption{\label{Fig3}
Patterns of the mean temperature field $\meanT$ in the $yz$ plane obtained in the laboratory experiments for different temperature differences $\Delta T$
between the bottom and upper walls of the chamber: (a) $\Delta T= 29$ K (upper panel);
(b) $\Delta T= 44$ K (middle panel); (c) $\Delta T= 61$ K (bottom panel).
The coordinates are measured in cm and temperature is measured in K.
}
\end{figure}

Let us discuss the obtained results.
In Fig.~\ref{Fig3} we show the patterns of the mean temperature field $\meanT$ in the $yz$ plane obtained in the laboratory
experiments for different temperature differences $\Delta T$
between the bottom and upper walls of the chamber: $\Delta T= 29$ K (upper panel);
$\Delta T= 44$ K (middle panel); $\Delta T= 61$ K (bottom panel).
Figure~\ref{Fig3} indicates that there are two large-scale circulations in the experiments for $\Delta T= 29$ K and
$\Delta T= 44$ K, and one large-scale circulation for $\Delta T= 61$ K.
The Rayleigh number ${\rm Ra}=\alpha_\ast \, g \, L_z^3 \, \Delta T /(\nu \, \kappa)$
in the experiments varies in the interval ${\rm Ra}=(0.61 - 1.23) \times 10^8$
depending on the different temperature differences $\Delta T$
between the bottom and upper walls of the chamber, where $\alpha_\ast$ is the thermal
expansion coefficient and $L_z$ is the height of the chamber.

In the present study we have not performed measurements of the velocity field.
The velocity field in turbulent convection in a similar experimental setup
has been investigated in our previous study \cite{BEKR09}, where
we have determined the mean and turbulent velocity fields,
the integral scales of turbulence in the horizontal and vertical directions,
the rates of energy dissipation and production in the convective turbulence,
the dependencies of these turbulent parameters on the values of the temperature difference
between the bottom and top walls of the chamber. For instance, the integral turbulence scales
in the chamber with convective turbulence are
$\ell_x=\ell_z=3$ cm, $\ell_y=5$ cm, the characteristic vertical turbulent velocity is $u_z =5$ cm/s at $\Delta T= 60$ K that
depends on $\Delta T$ as $u_z \propto (\Delta T)^{1/2}$ \cite{BEKR09}.

\begin{figure}
\centering
\includegraphics[width=8.5cm]{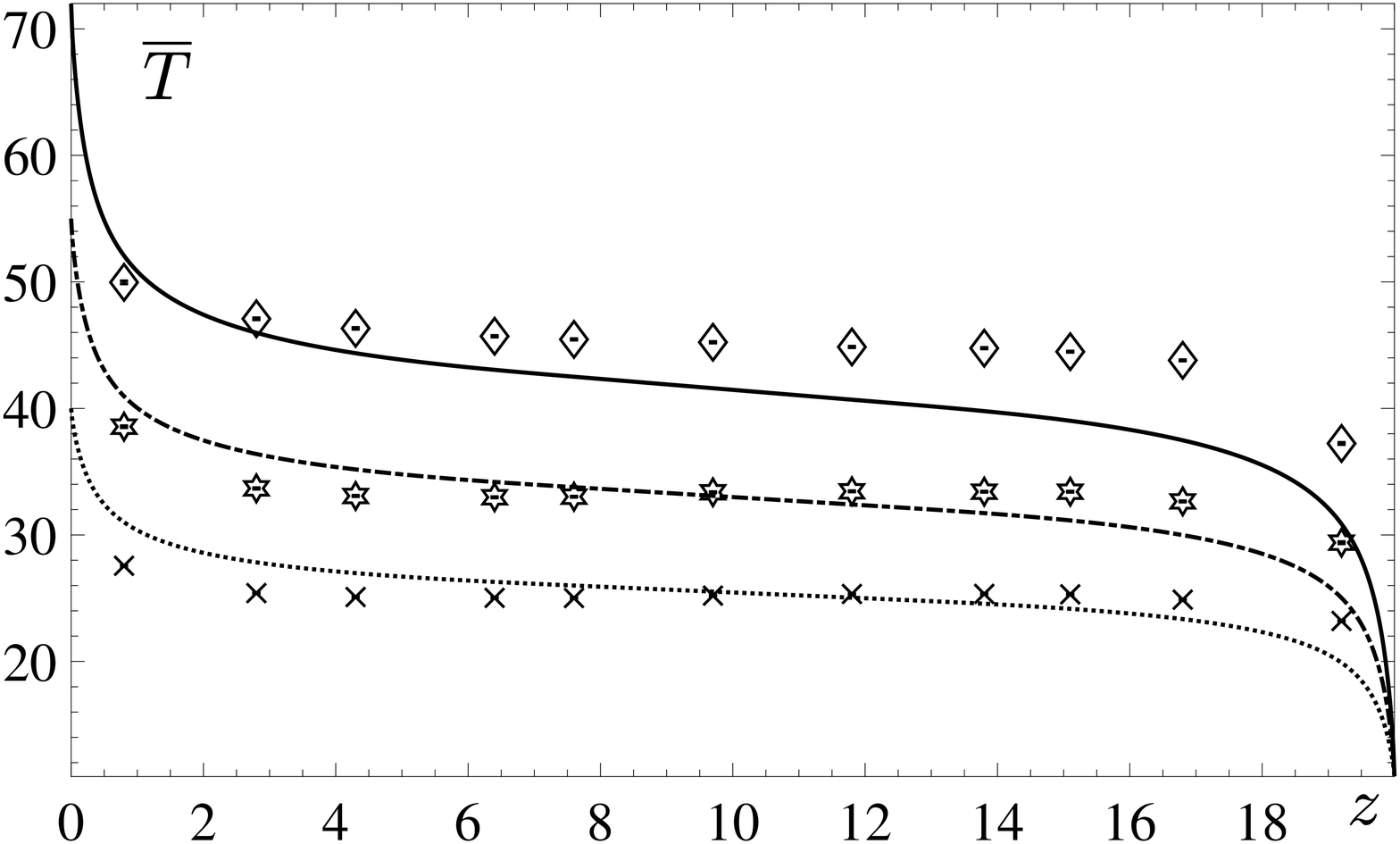}
\caption{\label{Fig4} Comparison of vertical profiles of the mean temperature $\meanT$ obtained
in the laboratory experiments for different temperature differences $\Delta T$
between the bottom and upper walls of the chamber: $\Delta T= 29$ K (slanting crosses);
$\Delta T= 44$ K (stars); $\Delta T= 61$ K (diamonds); and in the semi-analytical mean-field calculations:
$\Delta T= 29$ K (dotted line);
$\Delta T= 44$ K (dashed line); $\Delta T= 61$ K (solid line).
The coordinate $z$ is measured in cm and the mean temperature in C.
}
\end{figure}

\begin{figure}
\centering
\includegraphics[width=8.5cm]{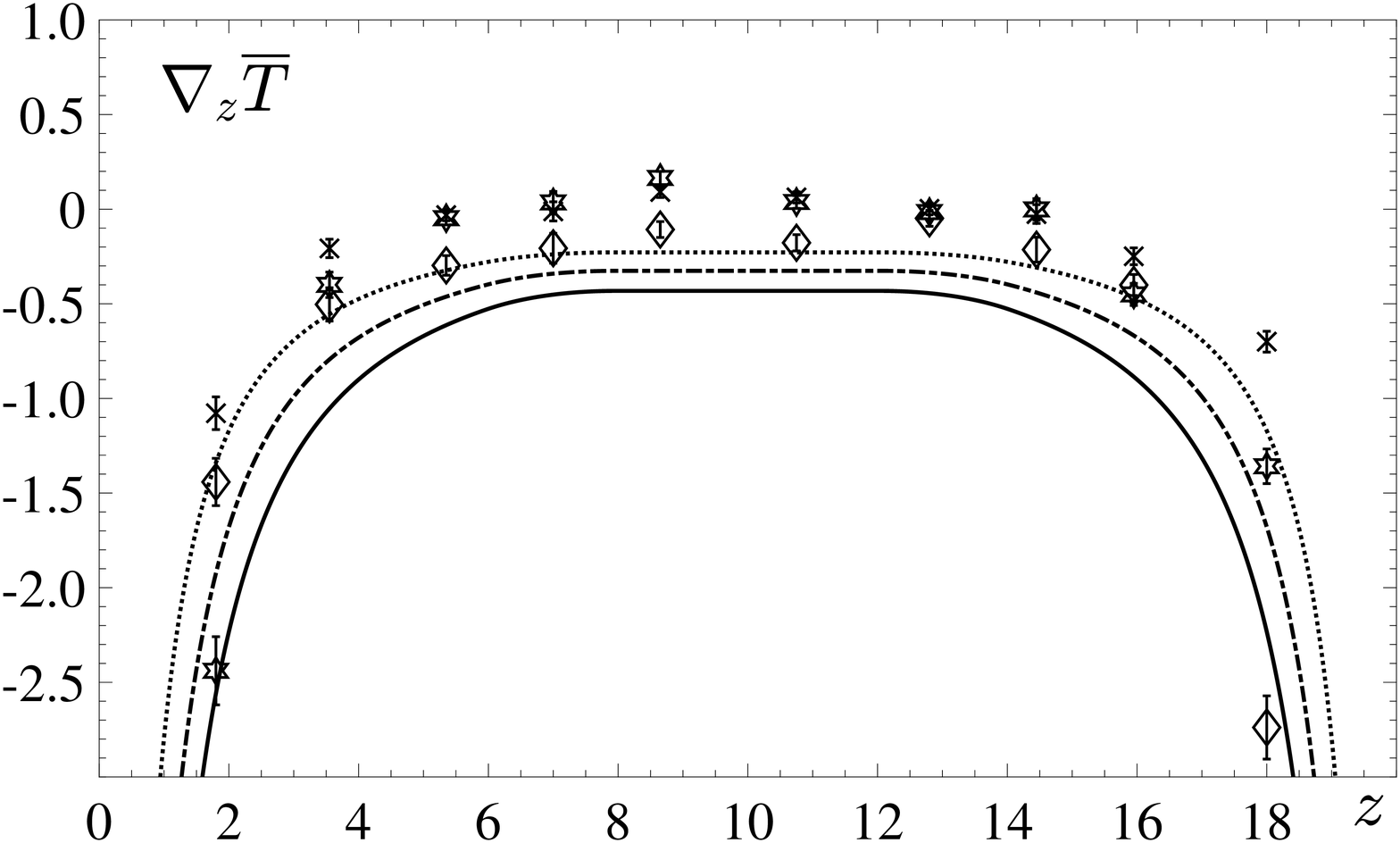}
\caption{\label{Fig5}
Comparison of vertical profiles of the vertical mean temperature gradient $\nabla_z \meanT$ obtained
in the laboratory experiments for different temperature differences $\Delta T$
between the bottom and upper walls of the chamber: $\Delta T= 29$ K (slanting crosses);
$\Delta T= 44$ K (stars); $\Delta T= 61$ K (diamonds); and in the semi-analytical mean-field calculations:
$\Delta T= 29$ K (dotted line); $\Delta T= 44$ K (dashed line); $\Delta T= 61$ K (solid line).
The coordinate $z$ is measured in cm and the mean temperature gradient $\nabla_z \meanT$ is measured in K/cm.
}
\end{figure}

In Figs.~\ref{Fig4} and~\ref{Fig5} we also plot vertical profiles of the mean temperature $\meanT$
and the vertical gradient  $\nabla_z \meanT$ of the mean temperature obtained in the laboratory experiments
for different temperature differences $\Delta T$
between the bottom and upper walls of the chamber.
These profiles of the mean temperature $\meanT$ are compared with the results of
the semi-analytical mean-field calculations described at the end of Section~\ref{sect3}.
Figures~\ref{Fig4} and~\ref{Fig5} show that there is a qualitative agreement
between the modelling and experimental results.
As usual, the strong temperature gradient is located near the walls
of the chamber, while inside the large-scale circulations the vertical gradient
$\nabla_z \meanT$ of the mean temperature is much weaker.
The main reason for the deviations of the modelling and experimental results
for the temperature field is that
in the one-dimensional semi-analytical calculations of the vertical mean temperature distribution,
we have not taken into account the mean velocity field of the large-scale circulation.
We have checked that a particular form of the decreasing smooth  function $\phi(z)$
that describes the spatial vertical profile of the integral turbulence scale in the region where
turbulent intensity strongly decreases,
weakly affects the mean temperature distribution.
Note also that we have not performed temperature measurements in the vicinity of the wall, but measure temperature
at the heat exchangers and in the core flow.

\begin{figure}
\centering
\includegraphics[width=8.5cm]{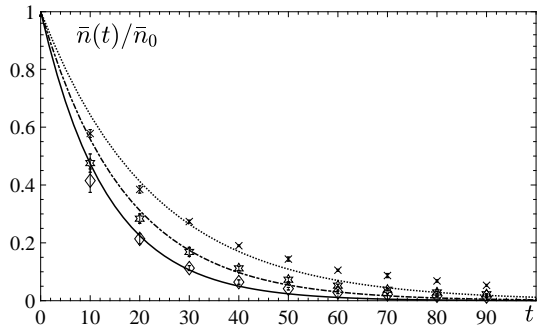}
\caption{\label{Fig6}
Comparison of the time evolution of the normalized mean number density of nanoparticles $\meanN(t)/\meanN_0$
obtained in the laboratory experiments for different temperature differences $\Delta T$
between the bottom and upper walls of the chamber: $\Delta T= 29$ K (slanting crosses);
$\Delta T= 44$ K (stars); $\Delta T= 61$ K (diamonds); and in the mean-field simulations:
$\Delta T= 29$ K (dotted line); $\Delta T= 44$ K (dashed line); $\Delta T= 61$ K (solid line).
Here $\meanN(t)=L_z^{-1} \, \int_0^{L_z} \meanN(t,z) \, dz$ and $\meanN_0=\meanN(t=0)$.
The time is measured in minutes.
}
\end{figure}

Measurements of number density of nanoparticles allow us to determine
the time evolution of the normalized mean number density $\meanN(t)/\meanN_0$  of nanoparticles obtained
in the laboratory experiments for different temperature differences $\Delta T$
between the bottom and upper walls of the chamber (see Fig.~\ref{Fig6}).
Here $\meanN(t)=L_z^{-1} \, \int_0^{L_z} \meanN(t,z) \, dz$ is the mean number density
of nanoparticles averaged over the vertical coordinate $z$ and $\meanN_0=\meanN(t=0)$.
Inspection of Fig.~\ref{Fig6} shows that
in the experiments with turbulent convection with different temperature difference between the bottom and top walls
of the chamber, the mean number density of nanoparticles decreases exponentially in time.
The characteristic decay time of the mean number density of nanoparticles varies
from 12.8 min for the temperature difference $\Delta T= 61$ K, to 16.3 min for $\Delta T= 44$ K and
to 24 min for $\Delta T= 29$ K.

We also perform comparison of the obtained experimental results with
the results of the mean-field numerical simulations of transport of nanoparticles
which accounts for molecular and turbulent effects for
the conditions pertinent to the laboratory experiments.
The numerical setup is described in Section~\ref{sect3}.
As follows from Fig.~\ref{Fig6}, the obtained numerical results
related to the time evolution of the normalized mean number  density of particles $\meanN(t)/\meanN_0$
are in an agreement with the results of the laboratory experiments.

\begin{figure}
\centering
\includegraphics[width=8.5cm]{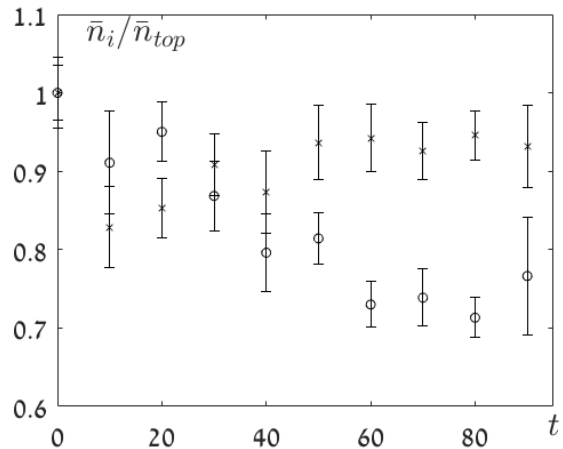}
\caption{\label{Fig7}
The time evolution of the ratios of the mean number densities of nanoparticles
$\meanN_{\rm bottom}/\meanN_{\rm top}$ (slanting crosses)
and $\meanN_{\rm middle}/\meanN_{\rm top}$ (circles)
obtained in the laboratory experiments for the temperature difference $\Delta T= 61$ K
between the bottom and top walls of the chamber, where
$\meanN_{\rm top}$,  $\meanN_i=\meanN_{\rm bottom}$
and $\meanN_i=\meanN_{\rm middle}$ are the mean number
densities of nanoparticles measured at the top, bottom and middle parts of the chamber, respectively.
The time is measured in minutes.
}
\end{figure}

In Fig.~\ref{Fig7} we show
the time evolution of the ratios of the mean number densities of nanoparticles
$\meanN_{\rm bottom}/\meanN_{\rm top}$ and $\meanN_{\rm middle}/\meanN_{\rm top}$
obtained in the laboratory experiments for the temperature difference $\Delta T= 61$ K
between the bottom and top walls of the chamber, where
$\meanN_{\rm top}$,  $\meanN_{\rm bottom}$
and $\meanN_{\rm middle}$ are the mean number
densities of nanoparticles measured in the vicinity of the top, bottom and middle parts of the chamber, respectively.
Figure~\ref{Fig7} demonstrates that the distribution
of the mean number density of nanoparticles is inhomogeneous.
In particular, the maximum mean number density of nanoparticles is located near the top cold wall of the chamber, i.e. for $t>0$,
the ratios $\meanN_{\rm bottom}/\meanN_{\rm top}$ and $\meanN_{\rm middle}/\meanN_{\rm top}$
are always less than 1.

Let us estimate the effective pumping velocity caused by turbulent thermal diffusion of nanoparticles,
the velocity due to the thermophoresis and the terminal fall velocity.
The effective pumping velocity caused by turbulent thermal diffusion of nanoparticles
having the diameter $d=70$ nm, is $7 \times 10^{-2}$ cm/s, where we use
Eqs.~(\ref{BBB8})--(\ref{BBB9}). We also take into account that
the value of the gradient of the mean temperature
measured in the core flow at $3 \leq z\leq 17$  cm is $\nabla_z \meanT = 0.4 $ K/ cm (see Fig.~\ref{Fig5}).
In this region the turbulent effects (turbulent thermal diffusion and turbulent diffusion)  are dominant.
Now we estimate the terminal fall velocity $V_g= \tau_{\rm p} \, g =3\times 10^{-5}$ cm/s,
where $\tau_{\rm p} =3\times 10^{-8}$ s is the Stokes time for nanoparticles
with the diameter $d=70$ nm.
The velocity due to the thermophoresis in the vicinity of the cold wall, is about $V^{\rm (tp)} =1.2\times 10^{-3}$ cm/s
at $z=1.6$ cm, where the gradient of the mean temperature is 3 K/ cm.

Note that several different mechanisms affect the particle transport:
the mean fluid velocity of the large-scale circulations,
turbulent diffusion, turbulent thermal diffusion, Brownian motions and
thermophoresis.
The mean fluid velocity of the large-scale circulations causes mixing
of nanoparticles in the chamber producing nearly homogeneous distribution
of nanoparticles in the chamber.
Since the mean fluid velocity in the downdrafts and updrafts inside the large-scale circulations
have opposite directions, the total vertical transport of nanoparticles
(averaged over the horizontal plane) due to the mean velocity is negligible and
it cannot cause the particle accumulation
in the vicinity of the cold wall of the chamber.
On the other hand, the effective pumping velocity
caused by turbulent thermal diffusion, is the vertical velocity directed
to the cold wall of the chamber (it is proportional to $- {\bf \nabla}\overline {T}$).
The effective pumping velocity results in an inhomogeneous distribution
of nanoparticles where  particles tend to be accumulated in the vicinity of the cold wall of the chamber.
However, turbulent thermal diffusion is a turbulent effect and it vanishes near the walls whereby
the intensity of turbulence tends to zero.
On the other hand, in the vicinity of the walls, the molecular effects (mainly the thermophoresis and
adhesion of nanoparticles) play a crucial role in the particle accumulation at the cold wall of the chamber.
Indeed, the thermophoretic velocity is proportional to $- {\bf \nabla}\overline {T}$,
and the mean temperature gradient is maximum in the vicinity of the walls.
Therefore, the thermophoretic velocity finally causes the trapping of particles by the cold wall of the chamber.

\section{Discussion and conclusions}
\label{sect6}

We have performed different sets of experiments with turbulent convection
to study turbulent transport of nanoparticles.
The temperature field has been measured with a temperature probe equipped with 11 E-thermocouples.
To determine the number density of nanoparticles, we used Condensation Particle Counter.
Nanoparticles of 70 nm in diameter are produced by Advanced Electrospray Aerosol Generator.
We measure the number density of nanoparticles as a function of time in 3 locations in vertical directions.
In the different experiments with turbulent convection with the temperature difference between the bottom and top walls of the chamber $\Delta T$ varying between $\Delta T= 29$ K to $\Delta T= 61$ K,
it has been shown that the mean number density of nanoparticles
decreases exponentially in time. For instance, the characteristic decay time of the mean number density of nanoparticles varies
from 12.8 min for the temperature difference between the bottom and top walls
of the chamber $\Delta T= 61$ K, to 16.3 min for $\Delta T= 44$ K and to 24 min for $\Delta T= 29$ K.
The main reasons for this effect are as follows.
\begin{itemize}
\item{
The large-scale circulations in convective turbulence cause mixing of nanoparticles in the chamber.}
\item{
The effective pumping velocity due to turbulent thermal diffusion in the core flow results in
the drift of nanoparticles to the cold wall of the chamber.}
\item{
The thermophoresis and adhesion of nanoparticles in the vicinity of the cold wall of the chamber where the mean temperature gradient is enough strong, result in trapping of nanoparticles by the cold wall.}
\end{itemize}

We have also performed one-dimensional mean-field numerical simulations of the evolution of the mean number density of nanoparticles which take into account turbulent and molecular effects.
The molecular effects include the Brownian diffusion of nanoparticles, the thermophoresis, and the adhesion of nanoparticles at the cold surfaces.
The turbulent effects are described by turbulent diffusion and turbulent thermal diffusion of nanoparticles.
These simulations allow us to determine the time-dependence
of the mean number density of nanoparticles and to find the characteristic decay time of the mean concentration of nanoparticles.
The obtained numerical results are in an agreement with the experimental results.

\begin{acknowledgements}
This research was supported in part by the Israel Ministry of Science and Technology (grant No. 3-16516).\\
\end{acknowledgements}

\medskip
{\bf DATA AVAILABILITY}
\medskip

The data that support the findings of this study are available from the corresponding author
upon reasonable request.

\end{document}